\documentclass{aa}

\usepackage{psfig}
\begin{document}

   \thesaurus{11     
              (11.01.2;  
               11.17.3;  
               )} 

   \title{Further Evidence that Some Quasars Originate in Nearby Galaxies: 
           NGC3628}

   \author{Eric Flesch\inst{1} \and  Halton Arp\inst{2}}

\institute{P.O. Box 411, Nelson, New Zealand \and Max-Planck-Institut 
f\"ur Astrophysik, 85740 Garching, Germany}
     
     \offprints{Eric Flesch, P.O. Box 411, Nelson, New Zealand;
      email: eric@flesch.org}

         \date{Received; accepted 1999}

        \maketitle

        \begin{abstract}

NGC3628 is a well-studied starburst/low level AGN galaxy in the Leo 
Triplet noted for its extensive outgassed plumes of neutral hydrogen. Catalogued 
QSOs are shown to be concentrated around NGC3628 and aligned with HI plumes.
The chance that the three nearest quasars would accidentally fall as close as
they do is $\leq 10^{-2}$. The nearest quasar has a redshift of  z = 2.15 and is at 
the tip of an
X-ray filament emerging along the minor axis plume. Location at this point has
an accidental probability of $\sim 2x10^{-4}$. In addition an unusual optical 
filament also points directly at this quasar.


\end{abstract}

\section{Introduction}
 
NGC3628 is a nearby edge-on Sbc peculiar galaxy which is undergoing 
major internal dynamic activity which is, however, shrouded from our view by a 
prominent dust lane.  For a comprehensive summary of the galaxy parameters 
and current observational status, see Cole, Mundell and Pedlar (1998).  
For a discussion of a possible AGN within NGC3628, see Yaqoob et al (1995).

A prominent feature of NGC3628 is the long HI plumes being outgassed from the 
galaxy in two directions.  The major plume to the ENE was first imaged by 
Kormendy and Bahcall (1974) as a long, straight optical jet . Later it was  
observed in HI with the Arecibo telescope by 
Haynes, Giovanelli and Roberts (1979) who also mapped out a weaker plume 
towards the south along the minor axis. The 
consensus explanation for the plume morphology is that it is consequential to a 
tidal encounter between NGC3628 and the nearby similar-sized spiral NGC3627, 
although the velocity profile and substructure of the plume caused Haynes et
al. to 
comment that "...the observational data somewhat strain the model parameters."  
It could also have been noted that comparable HI extensions were not
drawn out of adjoining galaxies with which NGC3628 was supposed to
have interacted. 

Haynes et al.'s definitive mapping of the plume morphology is replicated here in 
Fig. 1, minus the complete extension to the ENE, and without the 3K km/s 
contour about which Haynes et al. counseled caution.   The 
velocity profile of the major plume is that of steady flow velocity away from the 
galaxy out to beyond the left edge of Fig1. 

This excess hydrogen is originating from the central regions of NGC3628, either 
from starburst activity or from a dust-enshrouded AGN.  Fabbiano, Heckman and 
Keel (1990) concluded that the X-ray observations demonstrated "... collimated 
outflow from a starburst nucleus...", and Irwin and Sofue (1996) found expanding 
molecular shells of CO emanating from the nucleus.  NGC3628 is known to have 
a strong X-ray source in its core, and, unusually, a second strong X-ray source 
toward the east end of its disk.  Previous work by Arp (1997;1998) and others has
shown a 7.5 sigma association of quasars with Seyfert galaxies, attributed to
ejection from their active nuclei.  As NGC3628 is a prominent nearby galaxy 
which is 
so clearly seen to be expelling material from an active nucleus, it is suggestive 
that we should look there for an observational test of the nearby-quasar model.  
And as seen in Fig. 1, there are indeed surveyed quasars in the near vicinity of 
NGC3628, even in near propinquity to its active disk.  In this letter we will discuss 
evidence that these quasars are being vented out of NGC3628 along with the HI 
gas. 

\begin{figure*}
\mbox{\psfig{figure=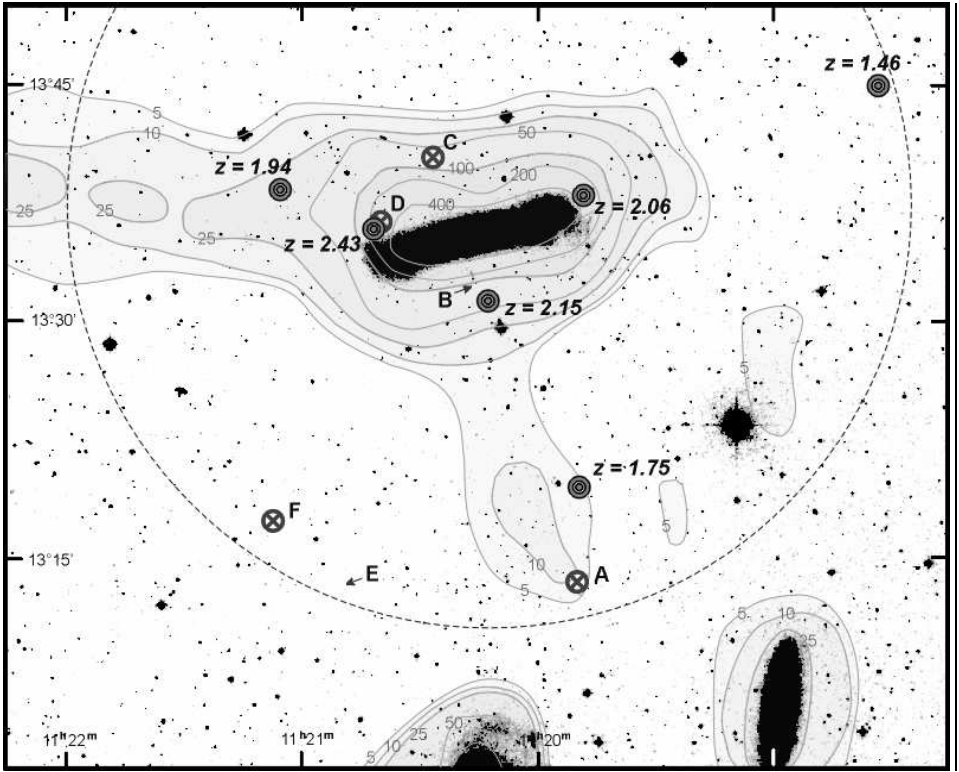,height=10.2cm}}
\rule{0.4pt}{10.2cm}
%\vspace{12cm}
%\includegraphics[width=\hsize]{ngc3628fig1.ps}
\hfill \parbox[b]{5.0cm}{\caption[]{Isophotal contours of neutral hydrogen (HI) 
coming from the 
starburst$/$AGN galaxy NGC3628 (from Haynes, Giovanelli and Roberts (1979)).  
Catalogued quasars are annotated with their redshift values, objects marked with 
a circled X are probable quasars (see table 1), and small 
arrows mark possible quasars.  The search field of Weedman (1985) is bounded 
by the dashed line. Objects surveyed by Weedman (1985) and by Dahlem et al 
(1996) are included in Fig 1.  Six catalogued QSOs are displayed, and six QSO 
candidates, the details of which are presented in table 1.}}%
\label{ngc3628fig1b}%
\end{figure*}

\section{Quasars and candidates near NGC3628}

Dahlem et al (1996) have ennumerated a population of X-ray emitting sources in 
the ROSAT-detected hot gaseous halo of NGC3628.  They found that this 
population density was over twice that of the wider background (=1.5 sigma 
deviation), but made no firm statistical statement due to the small numbers 
involved.  Dahlem et al. note that the emission properties of these sources point 
to their identification as either AGN (QSOs) or X-ray binaries, and as there is no 
physical model for X-ray binaries to be prevalent in the galaxy's halo, their 
identification as AGN (QSO) is indicated.  In the standard model all these QSOs 
are placed into the background, their greater density near NGC3628 
notwithstanding.  In the alternative model advanced by Arp (1987), G.R. Burbidge 
and others, 
most of these QSOs are local to the NGC3628 environment, which accounts 
simply for their observed overdensity near many such galaxies, see e.g., E.M. 
Burbidge (1999).

TABLE 1     \hspace{1.5cm}                     Objects in Fig 1.

\vbox{\tabskip=0pt \offinterlineskip
\def\tr{\noalign{\hrule}}
\halign to \hsize{\tabskip=1em minus 1em&
% plus2em&
#\hfill&#\hfill&#\hfill&\hfill#&#\hfill&#\hfill
\tabskip=0pt\cr
 ID& Survey\# & QSO &R.A.(2\rlap{000)~}&\hfill~~ Dec. &~~Notes\cr
\tr
  A &   Wee 49 &prob &11 19 49&~ 13 13 24 & two em lines \cr
  B &   Dahl 8&poss &11 20 15&~ 13 32 30 & X-ray, ext.  red \cr 
  C &   Dahl 13  &prob &11 20 26&~ 13 40 25&  X-ray,BSO \cr 
  D &   Dahl 15  &prob &11 20 40&~ 13 36 20&  X-ray,BSO,dbl\cr
  E &   Wee 53   &poss &11 20 51&~ 13 13 36&  one em line,brd \cr  
  F &   Wee 56   &prob &11 21 10&~ 13 17 54&  one em line\cr
}}

Thus we have 10 confirmed or probable QSOs in the surveyed area of figure 1, of 
which 8 are within the galaxy's plume contours. The average background density
which Weedman found for his 20 CFHT fields was 10/sq deg. for $2.0 \leq z <  2.5$
to m$_{4500}$ = 21 mag. For $1.75 \leq z < 2.5$ the density rises to 13.9/sq deg
+8/-6 which background density is plotted in Fig. 2. It is seen that the closest
quasars to NGC3628 reach a density of about 100/sq. deg. The chance of
finding the the two closest quasars at 19.7 and 19.9 mag. is only .002 (using
Poisson statistics). Of the three closest, the z = 2.43 is at 21.2 mag. and probably 
absorbed by the galaxy, but still makes the probability of chance
concentration even smaller. Also, this is based on an over-prediction of 
background numbers because, as we have seen the Weedman survey set on 
some very active galaxies and then only used plates with the most quasars 
below m$_{4500}$ = 20 mag. 

\begin{figure}
%\vspace{4cm}
\mbox{\psfig{figure=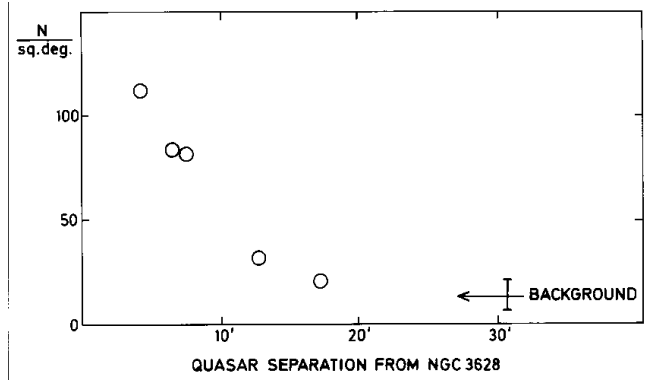,height=5cm}}
%   \vspace{11.0cm}
    \caption[]{The density of catalogued quasars at each distance  from NGC3628. 
is plotted. The arrow marks the upper limit to average 
background density of $1.75 \leq z \leq 2.5$ quasars. The area blocked by 
NGC3628 has been excluded.}

     \label{ngc3628fig2}
\end{figure}

\section{Quasars in the galactic disk of NGC3628}

 In Fig.1 the z = 2.43 quasar is the X-ray source Dahlem \#16. Only 37
arcsec away is the X-ray source Dahlem \#15 which we have identified with a
blue stellar object (BSO) of similar apparent magnitude. Apparently this is a
double quasar of which the second component remains to be observed
spectroscopically. Optical identification of these objects was enabled from the
coordinates of Read et al. (1997) and interchange of the published declinations of
Dahlem \#15 and \#16 (M. Dahlem, private communication).

     The quasars at both  the east and west end of the disk appear to be
associated with strong disturbances in NGC3628 at these points. Luminous
features and dust features point in the general direction of the quasars and there 
are hints of filaments and perturbations which, when explored with deeper, higher
resolution images from larger telescopes, may link these quasars with the
general eruption of material in these regions.

\section{Quasars in the HI Plumes}

Fig.1 shows alignment between the HI plumes of NGC3628 and nearby quasars.  
One quasar (z=1.94) lies at the base of the main ENE plume, coincident onto the 
optical jet imaged by Kormendy and Bahcall (1974).  Another quasar (z=1.75) 
lies in 
the southern plume coincident with a thickening of the plume, and the probable 
quasar Wee49 (Weedman 1985) lies, very interestingly, right at the tip of the 
southern plume.  Thus these quasars are not only aligned with the southern 
plume, but positioned at contour nodes.  This is strongly indicative of physical 
association, and implies that these quasars and HI plumes have come out of 
NGC3628 in the same physical process.

\section{X-ray Ejection from the Nucleus of NGC3628}

As referenced previously, the first X-ray observations established collimated 
outflow along the minor axis of NGC 3628.  The later ROSAT observations 
confirmed this result and established narrow filaments and point sources 
extending outward from the nucleus (Dahlem et al, 1996; Read, Ponman and 
Strickland 1997).  Fig. 3 here, the best resolution PSPC X-ray map, shows a 
narrow filament coming from the bright X-ray nucleus continuously out to end on 
the z=2.15 quasar. 

     If we adopt the last X-ray source as the end of the filament, then the quasar
falls essentially on its tip. The accuracy of the superposition is obtained by taking
the X-ray position of Dahl\#7 and differencing it with that of the APM position of
the quasar. That yields a displacement of $\sim$20 arcsec and a probability of
accidental superposition of $\sim10^{-3}$. However, identifying optically the
stronger Dahlem sources gives $\sim$12 arcsec systematic correction for the X-ray
positions. That yields a coincidence of 8 arcsec, about the accuracy of PSPC
identifications, and a probability of accidental coincidence of $2x10^{-4}$.    

\begin{figure}
%\vspace{5cm}
\mbox{\psfig{figure=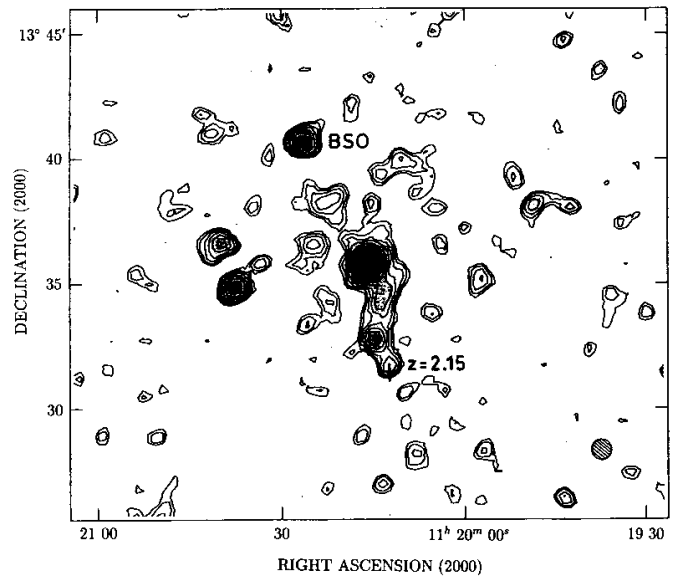,height=7.8cm}}
%   \vspace{9.0cm}
    \caption[]{The 0.75 keV band map of NGC3628 from Dahlem et al. (1996).  The 
optical position of the z = 2.15 quasar is marked by a plus at the tip of the 
filament coming out along the minor axis to the SSW. BSO marks a blue stellar 
object identified with the X-ray source Dahlem \#13.}
     \label{ngc3628fig3}
\end{figure}

The quasar, as marked in Fig. 3, is an X-ray source and the 
next source in toward the nucleus also is a point X-ray source which should be 
investigated spectroscopically.  Optically this latter source appears slightly 
extended so it should be particularly interesting.
At about an equal distance on the other side of the NGC3628 nucleus, along the 
N minor axis, is a brighter X-ray source which is identified with an E = 18.6 mag., 
blue stellar object (BSO), shown as object C on Fig. 1.  This is almost certainly a 
quasar and when confirmed will form a pair across NGC3628 like so many other 
pairs which have now been found across active galaxies (see Arp 1997; 1998).

\section{The Optical Filament Leading to the z = 2.15 Quasar}

Fig. 4 shows a processed portion of the Palomar Schmidt E plate (POSSII).  A 
short, narrow optical filament emerges along the minor axis nearly coincident 
with the X-ray filament pictured in Fig. 3.  The simplest model would be that this 
is gas from the interior regions of NGC3628 which has been entrained in the 
ejection of whatever is coming, or has come out, along the line of the X-ray 
ejection.  The z = 2.15 QSO (Wee 51) is identified with Dahlem et al. (1996) 
object \#7, also identified 
by Read, Ponman and Strickland (1997) as their object \#2.  Optical identification
in Fig.1 of the optically-extended, point X-ray object Dahlem et al. (1996) \#8,  is 
enabled by positional projection from the QSO using both papers' co-ordinates.  

\begin{figure}
%\vspace{7.5cm}
\mbox{\psfig{figure=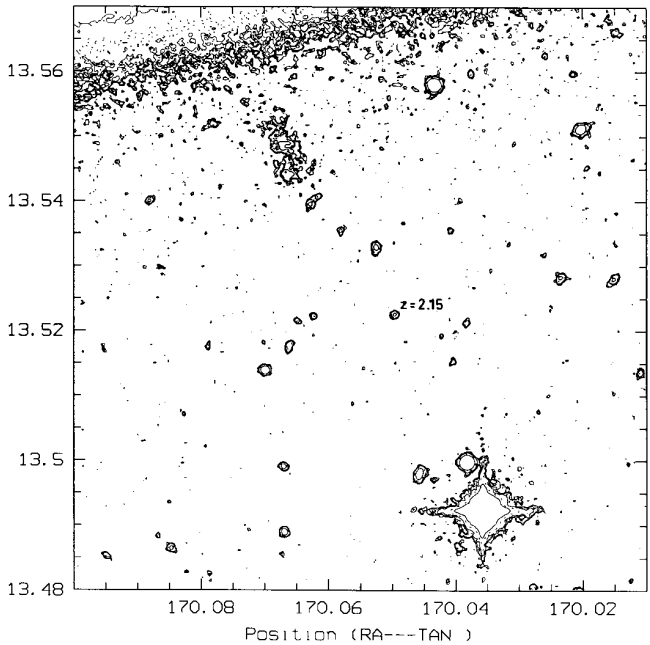,height=8.8cm}}
%   \vspace{9.0cm}
    \caption[]{An optical filament coming out along the minor axis of NGC3628 
(edge of galaxy at upper left).  The knots leading to the z = 2.15 quasar are blue 
and slightly extended and lie in the X-ray filament pictured in Fig. 3.  Image from 
processed, POSSII Schmidt plate.}

     \label{ngc3628fig4}
\end{figure}

There is an excellent precedent for this kind of an optical filament in the giant 
radio galaxy CenA (NGC5128).  There the optical filament coincides with the 
direction of both the radio jet and the X-ray jet further in the interior.  The optical 
filament consists of young stars and HII region-like emission lines, similar to a 
star forming arm in a spiral galaxy (Blanco et al. 1975; Arp 1986).  Hence we 
would expect a low redshift emission line spectrum for the optical filament in 
NGC3628.  But in NGC3628 the z = 2.15 quasar is at the tip of the filament and 
there is an X-ray point source in the filament as well as several blue and/or 
slightly extended optical objects.  We should be able in this case to investigate 
the mechanics of the entrainment and excitation if better resolution plates and 
spectra are obtained.

In general, the narrowness of the filament requires that whatever is ejected in the 
X-ray jet (and in the case of Cen A the coincident radio jet) must be quite small.  
There would seem to be no candidates other than quasars, which are generally 
X-ray and radio sources and exhibit similar spectra to the active nuclei which are 
actually ejecting the material.  Figures 3 and 4 would then represent a fortuitous 
moment when the quasar is just passing out beyond the filament.

\section{Redshift vs separation from galaxy}

The redshifts of the quasars in Fig. 1 decrease as their projected
angular distances from NGC3628 increase. This has been found previously for 
galaxy-quasar associations but in the NGC3628 case the z vs lnr plot
decreases less steeply than for quasars coming out only along galaxy minor
axes (Arp 1999, Fig. 3). This may be because we are dealing with quasars mostly
around z = 2 here, or it may mean that the quasars here are interacting with the
material of the galaxy and being slowed or captured from their less inhibited
escape along minor axes.

\section{Summary}

In a completely searched area around NGC3628 the known quasars are 
concentrated to the position of the starburst/ AGN galaxy.  Two quasars and a 
probable quasar are situated on the rim of the galaxy disk.  The galaxy has two 
prominent plumes of hydrogen gas and quasars are well-aligned with key 
points on the plume contours.  On the minor axis of NGC3628 there are a quasar 
and a probable quasar on opposite sides of the galaxy, which is a prototypical 
configuration.  {\it Perhaps most striking of all, a narrow X-ray and optical
filament points along the minor axis of NGC3628, directly to the closest quasar.} 

    We believe the improbability of finding quasars so close to NGC3628, including
one of them linked directly to the nucleus, combined with the improbablility of
finding the galaxy to be so actively ejecting associated plumes of gas, optical
and X-ray material presents a key confirmation of the origin of quasars.  A search 
for further quasars located within the solid angle of the bright disk of 
NGC3628, spectroscopic identification of the surrounding X-ray
BSO's and analysis of the optical filament would give further insight into the
physical mechanisms of their origin.

\end{document}